\documentclass[aps,prb,twocolumn]{revtex4}
\usepackage{graphicx}
\usepackage{epsfig}
\usepackage{amsmath}
\usepackage{amssymb}
\usepackage{amsfonts}%
\begin{document} 
\title{Spin relaxation in narrow wires of a  
two-dimensional electron gas}  
\author{Peter Schwab}
\author{Michael Dzierzawa}
\author{Cosimo Gorini}
\affiliation{Institut f\"ur Physik, Universit\"at Augsburg, 86135 Augsburg, Germany}
\author{Roberto Raimondi}
\affiliation{Dipartimento di Fisica "E. Amaldi", Universit\`a di Roma Tre, Via della Vasca Navale 84, 00146 Roma, Italy}

\begin{abstract} 
How does an initially homogeneous
spin-polarization in a confined two-dimensional electron gas with Rashba
spin-orbit coupling evolve in time? How does the relaxation time
depend on system size? We study these questions for systems of a size that
is much larger than the Fermi wavelength, but comparable and even
shorter than the spin relaxation length. 
Depending on the
confinement spin-relaxation may become faster or slower than in the
bulk. An initially homogeneously polarized spin system evolves into a
spiral pattern.

\end{abstract} 
\pacs{PACS numbers: } 

\date{\today} 
\maketitle 
\section{Introduction}
Long spin relaxation times are desirable for the operation of
spintronic devices. 
In confined systems like quantum dots with a discrete level spectrum
an initial electron spin polarization will not decay exponentially,
unless the spins are coupled to
extra degrees of freedom like phonons or nuclear spins.
\cite{khaetskii2000,merkulov2002}
For larger or open systems, where the energy spectrum can be considered
continuous, exponential decay is possible by purely elastic scattering
which is then the dominant process at low enough
temperatures.

In this article we reconsider spin-relaxation
in a two-dimensional electron gas  confined to a
narrow channel, concentrating on elastic scattering processes.
Our study has been motivated by a recent experiment, where 
the size dependence of the spin relaxation time of an $n$-InGaAs wire has been measured
via Faraday rotation spectroscopy.\cite{holleitner2006} 
The size dependence sets in when the wire width is of the order of several bulk spin relaxation lengths.
The observed behavior is non-monotonic with an initial increase followed by a sharp decrease at the
smallest wire widths. The maximum spin relaxation time occurs when the width is of the order of the bulk
spin relaxation length.
Remarkably, while the tendency to suppress spin-relaxation in  confined systems has been predicted in a number of
theoretical works, 
\cite{malshukov2000,halperin2001,aleiner2001,kisilev2000,zaitsev2005,chang2004}
there has been no anticipation of the increase of the spin relaxation observed at the smallest widths.
In this article we show that spin active  boundaries, not considered in the previous theoretical analysis,
dramatically change  the size dependence of the spin relaxation time in the small width limit and  provide
a useful point of view as far as the interpretation of the experiment is concerned.

The rest of the paper is organised as follows. 
In Sec.\ II we recall the method of quasiclassical Green functions 
and point out when a diffusion equation approach is possible. 
In Sec.\ III we consider the spin dynamics in the bulk,
whereas in Sec.\ IV we turn to spin relaxation in finite systems. A
summary is given in Sec.\ V.
\section{The basic equations}
We start from the Hamiltonian of a two-dimensional electron gas
\begin{equation} \label{eq1}
H = \frac{p^2}{2m} + {\bf b} \cdot \boldsymbol \sigma +V({\bf x})
,\end{equation}
where ${\bf b }$
is the internal magnetic field due to the spin-orbit
coupling, ${\boldsymbol \sigma}$ is the vector of Pauli matrices and
$V({\bf x})$ is a delta-correlated random potential due to impurity scattering.
For the internal field we concentrate on the Rashba model,
${\bf b} = \alpha \, {\bf p} \times {\bf e}_z $,
where the spin-orbit field arises due to a structural inversion
asymmetry.  However, our treatment can be applied also to the case
of bulk inversion asymmetry, where  there is in addition the
Dresselhaus term\cite{dresselhaus1955}  which contributes to the two-dimensional effective Hamiltonian with 
${\bf b} = \beta (p_x, -p_y ) + \gamma( p_x p_y^2, -p_y p_x^2)$, or 
to the case of a  two-dimensional hole gas, where  the Rashba spin-orbit field 
has  a cubic dependence on momentum.
\cite{winkler2000}

To study the spin dynamics we rely on the method 
of the quasiclassical Green function.
\cite{rammer1986,schwab2003} 
The latter solves the Eilenberger equation ($\hbar =1$),
\begin{equation} \label{eq2}
\partial_t g + { \bf v }_F\cdot \nabla g +
{\rm i } [{\bf b} \cdot {\boldsymbol \sigma}, g ] = 
- \frac{1}{\tau} \left(
g - \langle g \rangle
\right)
,\end{equation}
where $\tau$ is the elastic scattering time, 
arising by the adoption of the standard self-consistent Born approximation for elastic impurity scattering.
${\bf v}_F$ is the Fermi velocity and  $\langle  \dots \rangle $ denotes the angular average over the Fermi
surface.
Eq.~(\ref{eq2})  is valid in the limit when both the spin-orbit energy and $\tau^{-1}$ are  small
compared to the Fermi energy.
\cite{raimondi2006}
Besides the space and time dependence implied by Eq.~(\ref{eq2}),
the quasiclassical Green function $g=g_{ss'}({\epsilon},  \hat{\bf   v}_F; {\bf x}, t) $
is a matrix in spin space and is a function of  
energy $\epsilon$ (measured with respect to the Fermi surface) and 
direction on the Fermi surface, $\hat {\bf v}_F$. 
\cite{keldysh}
In the absence of external potentials and magnetic fields 
the particle and spin density are obtained by summing over energy, angle and spin 
\begin{eqnarray}
\rho({\bf x}, t)   &  = &  - \frac{1}{4}N_0 \int \!  {\rm d }\epsilon 
\sum_s \langle  g_{ss}( \epsilon, \hat {\bf v}_F ; {\bf x}, t )\rangle,   \\
{\bf s}({\bf x}, t) & =  & - \frac{1}{4}N_0 \int{\rm d }\epsilon
\sum_{ss'} 
\frac{1}{2} 
{\boldsymbol \sigma}_{ss'} 
\langle  g_{s's}(\epsilon, \hat{ \bf v}_F  , {\bf x}, t ) \rangle
.\end{eqnarray}

When the spin-orbit energy is small compared to the
scattering rate 
the spin-dynamics becomes diffusive in the low frequency, long
wavelength limit.
In this limit, all the harmonics characterizing the angle dependence
of $g$ can be expressed conveniently in terms of the $s$-wave
component.  
As a result the problem simplifies considerably and one obtains, 
for the angle averaged Green function, $\langle g \rangle$, a diffusion equation, 
which, in the presence of a general spin-orbit field, acquires the form
given, for instance, in Ref.\ \onlinecite{malshukov2005}. For the
Rashba Hamiltonian in particular the equations read
\begin{eqnarray} \label{eq5}
\left( \partial_t -  D \partial_{\bf x}^2  \right) \rho & = & 0 \\
\label{eq6}
\left( \partial_t -  D \partial_{\bf x}^2 \right) s_x   & = &   -
\frac{1}{\tau_s}   s_x
+2C \partial_x s_z  \\
\label{eq7}
\left( \partial_t -  D \partial_{\bf x}^2 \right) s_y    &= &
-\frac{1}{\tau_s}   s_y 
+2C \partial_y s_z   \\
\label{eq8}
\left( \partial_t -  D \partial_{\bf x}^2 \right) s_z &= &  -
\frac{2}{\tau_s}   s_z
- 2C\partial_x s_x  - 2 C \partial_y s_y  
,\end{eqnarray}
with $D=\frac{1}{2}v_F^2 \tau$, $\tau_s = \tau/[2(\alpha p_F \tau
)^2]$, and $C= v_F \alpha p_F \tau$. 
Since here we focus on the spin dynamics, we limit ourselves to  terms to leading order in the
parameter $\alpha/v_F$, therefore neglecting spin-charge coupling
\cite{burkov2004,mishchenko2004,malshukov2005,raimondi2006}.

%
%
%
\section{Spin dynamics in the bulk}
The spin-dynamics is particularly simple for a spatially homogeneous 
spin density. From the diffusion equation one observes that the
different spin directions decouple and the spin polarization along the ${\bf e}_{x,y}$ or ${\bf e}_z$ axis
relaxes with the time constant $\tau_s$ or $\tau_s/2$,
respectively. This type of spin relaxation, known as the D'yakonov-Perel'
mechanism, \cite{dyakonov1971} is easily understood:
the electron spin precesses around the internal field ${\bf b}$.
Scattering from an impurity changes the direction of the internal
field, and thus randomizes the spin precession.
If the precession angle, $\delta \Phi \sim |{\bf b} | \tau $, between collisions is
small, the spin dynamics resembles a diffusion. The total precession
angle after $N$ scattering events is of the order $\sqrt{N } \delta
\Phi$, from which the spin relaxation rate is estimated  of the order 
$\tau_s \sim  \tau /\delta \Phi^2 \sim \tau/ (\alpha p_F \tau )^2 $.

In general we cannot rely on the diffusion equation but have
to solve Eq.~(\ref{eq2}).
In the spatially homogeneous case the task simplifies, since again 
different spin directions do not couple. As a result, the spin dynamics is  
described by \cite{raimondi2006}
\begin{eqnarray}
\left[  L(L^2 +a^2) -L^2 - \frac{1}{2} a^2
\right] s_{x,y} &= & 0 \\
\left[ L( L^2 + a^2) -L^2  \right] s_{z  } &= & 0 
\end{eqnarray}
with $L = 1 + \tau \partial_t$ and $a= 2 \alpha p_F \tau$.
Apparently the full description of the spin dynamics now requires
three time constants for each component. 
The slowest decaying component goes as
$s_{i}(t) \sim \exp(-\gamma_i t) $ with 
\begin{eqnarray}
\gamma_{x,y } & =  &1/\tau_s  \quad \quad (a \ll 1) \\
\gamma_{x,y}  &  = &1/2 \tau  \quad \quad (a \gg 1) \\
\gamma_z      & =  &\frac{1}{2\tau} - \frac{1}{2\tau } \sqrt{1-4a^2} 
,\end{eqnarray}
i.e.\ for a clean system all components decay on the time scale of the scattering time
$\tau$.

For conventional diffusion the mode with the longest lifetime is
homogeneous in space. Here we encounter a different situation, due to the
coupling of the various spin components.
From the diffusion equation, for instance, the modes with the longest lifetime
are found to form an elliptically modulated spin spiral,
\begin{equation}
{\bf s}({\bf x},t ) \propto  [ {\bf e}_q \cos({\bf q} \cdot {\bf x })
 + A {\bf e}_z \sin({\bf q} \cdot {\bf x}) ] \exp(- \gamma_0 t )
,\end{equation}
where the vector ${\bf q}$ lives in the $x$-$y$-plane with
$Dq^2 = 15 / 16 \tau_s$, $A= 3/\sqrt{15}$, and $\gamma_0 = 7/16 \tau_s$.
These long-living modes have been investigated  in
[\onlinecite{froltsov2001}]
and in [\onlinecite{pershin2005}].
%
%
%
\section{Spin dynamics in finite systems}
For finite systems the equations have to be supplemented by boundary
conditions. \cite{zaitsev1984,shelankov1985,millis1988}
For specular scattering where an in-going 
trajectory is scattered into one outgoing direction 
the boundary condition for the Eilenberger equation reads \cite{millis1988}
\begin{equation}\label{eq15}
g_{ss'}^{\rm out }  =  \sum_{s_1 s_1'} S_{ss_1}
g_{s_1 s_1'}^{\rm in } S^+_{s_1's'}
.\end{equation}
The unitary matrix $S$ describes the surface scattering. 
By decomposing  the Green function in charge and spin components,
$ g_{ss'} = g_0 \delta_{ss'} + {\bf g} \cdot {\boldsymbol \sigma}_{ss'}$,
the boundary condition is equivalently expressed as
\begin{equation} \label{eq16}
g_0^{\rm out}       =  g_0^{\rm in}, \quad  {\bf g }^{\rm out}  =
 R  \, {\bf g}^{\rm in}
\end{equation}
with an orthogonal matrix $ R$ that rotates the spin at the
surface. 
Charge conservation implies that no current flows through the
boundary,
\begin{equation}
   \langle  {\bf n} \cdot {\bf v}_F \, g_0 \rangle \propto  {\bf n} \cdot {\bf j}_c = 0
,\end{equation}
where ${\bf n}$ is a vector normal to the boundary.
For a spin conserving boundary ($R=1$) also all components 
$\alpha$ of the 
spin current perpendicular to the surface are zero, 
\begin{equation} \label{eq18}
    \langle  {\bf n} \cdot {\bf v}_F \, {  g}_\alpha \rangle \propto  {\bf n} \cdot
{\bf j}_\alpha = 0 
.\end{equation} 
For the general case, $R \ne 1$, Eq.~(\ref{eq18}) is not valid.

In the following we consider two types of boundary conditions.
First we consider a spin-conserving boundary, where
\begin{equation}\label{hardwall}
| {\bf k}_{\rm in }  \, s \rangle \to | {\bf k }_{\rm out} \,  s \rangle 
,\end{equation}
such that $S$ and $ R$ are unit matrices. 
As a second example we will consider a boundary
that scatters adiabatically, i.e.,\ an incoming wave in an eigenstate 
$|{\bf k}_{\rm in }  \, \pm \rangle$ of the Hamiltonian (\ref{eq1}) --
not including disorder -- is scattered into the same band,
\begin{equation}\label{softwall}
| {\bf k}_{\rm in }  \, \pm \rangle \to | {\bf k }_{\rm out} \,  \pm
\rangle
,\end{equation}
as it is expected for a smooth confining potential.
\cite{silvestrov2005,chen2005}
Since the eigenstates in the presence of the field ${\bf b}$
are spin-polarized parallel to ${\bf b}$,
the adiabatic boundary implies a rotation of the spin polarization,
$S=e^{{\rm i}\sigma_z \varphi}$,
where $\varphi$ is the angle between the electron momentum and the normal
of the interface.

Before presenting our numerical results for the spin relaxation obtained
by solving Eq.\ (\ref{eq2}),
it is useful to derive the boundary conditions in the diffusive regime.
The idea is to solve the Eilenberger equation near the boundary
assuming that the angular average of $g$ 
varies only slowly on the scale of the mean free path, $l$.
This is justified when the spin relaxation length $L_s$ is much longer
than $l$.
For directions pointing into the boundary one can then use the
expansion valid in the bulk,
\begin{equation} \label{eq17}
{\bf g}^{\rm in} = \langle {\bf  g } \rangle 
 - \tau {\bf v}_F \cdot \nabla \,  \langle {\bf g} \rangle
 +2 \tau  {\bf b} \times  \langle {\bf  g } \rangle 
.\end{equation}
Combining (\ref{eq16}) and (\ref{eq17}) yields a relation that connects linearly the 
three spin components of the angular averaged Green function,
$ \langle {\bf g} \rangle $, and its spatial derivatives.
The resulting form of the boundary condition depends on
the rotation matrix $ R(\varphi)$,
\begin{equation}\label{eq20}
\int_{\varphi_{\rm in } } \frac{ {\rm d} { \varphi} }{2 \pi }
[1+  R ] \left[  \langle  {\bf g } \rangle
 - \tau {\bf v}_F \cdot \nabla \, \langle {\bf  g }\rangle
+  2  \tau {\bf b} \times \langle {\bf g } \rangle \right]        =
  \langle  {\bf  g }\rangle.
\end{equation}
For the Rashba model, using the expression for the charge and the spin current in the diffusive limit, we get 
for spin-conserving boundary conditions, \cite{malshukov2000,galitski2006}
\begin{eqnarray} \label{eq22}
   -D \partial_x  s_x    - C s_z 
    =    {\bf n} \cdot {\bf j}_x      & = &0 ,\\
   \label{eq23}
   -D \partial_x  s_y   =      {\bf n} \cdot
   {\bf j}_y  & = & 0 , \\
   \label{eq24}
   - D  \partial_x  s_z   + C  s_x    =  {\bf n} \cdot
   {\bf j}_z & = &0
,\end{eqnarray}
where  ${\bf n}$ is in the $x$-direction. 
For adiabatic boundary conditions, in contrast, angular averaging of
Eq.\ (\ref{eq20}) yields that 
\begin{equation} \label{eq26a}
s_x=0 \, \, \text{and}  \, \, s_y=0,
\end{equation}
while the $z$-component of the
spin is still conserved and therefore Eq.\ (\ref{eq24}) remains valid.

\begin{figure}
\includegraphics[width=0.42\textwidth]{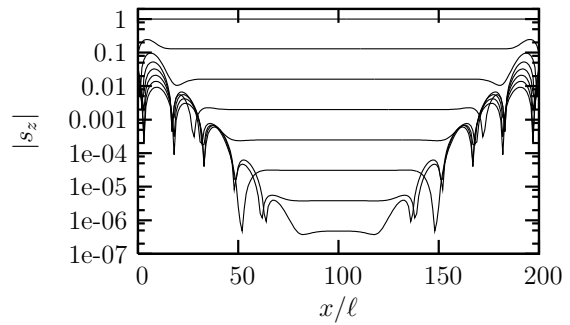}
\caption{\label{fig1}Time evolution of the spin polarization in
a wide channel ($L=200 \, l\approx  40 \, L_s$, 
$\alpha p_F \tau = 0.1$).
The curves from top to bottom correspond to different times, with 
$\Delta t = 50 \tau \approx  \tau_s$. $s_z$ changes sign at various
positions where a
steep
drop of $|s_z|$ is visible in the figure.
}
\end{figure}
Fig.\ \ref{fig1} shows the time evolution of the spin profile
for a long wire of width $L= 200 \, l$,
where $l=v_F \tau $ is the elastic mean free path.
Initially the spin was homogeneously polarized in $z$-direction.
The results were obtained from the Eilenberger equation with $\alpha
p_F \tau = 0.1$, and the conserving boundary condition, $S_{ss'}=\delta_{ss'}$.
Inside the wire one observes a homogeneous decay of the spin
polarization, with the time constant $\tau_s/2$. At the boundaries long
living modes show up which dominate the spin profile in the long time limit.

\begin{figure}
\includegraphics[width=0.42\textwidth]{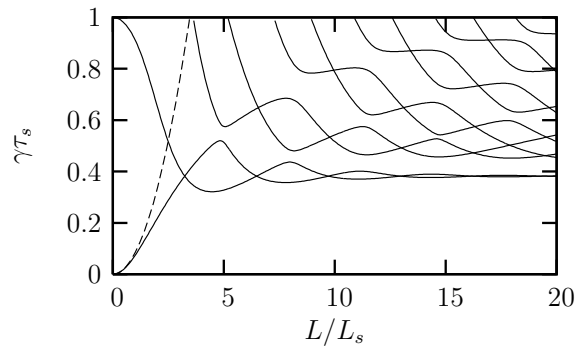}
\caption{\label{fig2}
Lowest eigenvalues of the spin-diffusion operator for the
Rashba model and conserving boundary conditions,
Eqs.~(\ref{eq22})--(\ref{eq24}). Modes with $\gamma <7/16 \tau_s$
have a complex wave vector, and can therefore exist only at the edges
of the wire. The dashed curve is $\gamma \tau_s = (L/L_s)^2/12$ obtained
in  Ref.\ \onlinecite{malshukov2000} for very narrow wires.}
\end{figure}
For further investigation of these modes we write the spin diffusion
equation, Eqs.~(\ref{eq6})--(\ref{eq8}), as $\partial_t {\bf s} + \hat
\gamma {\bf s} = 0$, and determine the eigenvalues and eigenmodes
of the operator $\hat \gamma$.
The eigenmodes are superpositions of plane waves. 
The low frequency spectrum of $\hat \gamma$ is shown in Fig.\
\ref{fig2} as a
function of the wire width.
Recall that the smallest eigenvalue for a bulk system is 
$\gamma_0 = 7/16 \tau_s$.
The modes with smaller decay rate have a complex wave vector 
and are thus localized near the edges of the wire.
For a wide system we find 
a continuum of eigenvalues above $\gamma_0$, and two localized modes
at $\gamma \approx 0.382 /\tau_s$.

For a narrow wire most strikingly one eigenvalue goes to zero with
decreasing width, asymptotically as 
\begin{equation}
\gamma \tau_s  \simeq  \frac{1}{12} 
\left(     \frac{L}{L_s} \right)^2. \end{equation}
This corresponds to the suppression of spin-relaxation in small systems reported 
earlier by other authors.
\cite{malshukov2000,kisilev2000,halperin2001,aleiner2001,chang2004,zaitsev2005}
This effect can be traced back to the specific form of
the spin-orbit field in the Rashba Hamiltonian -- being proportional
to the velocity. \cite{halperin2001}
Here we formulate the argument for a system including Rashba and also 
linear Dresselhaus term within the spin-diffusion equation approach.
For a spin profile that is homogeneous in the
$y$-direction the angular averaged Eilenberger equation, Eq.\ (\ref{eq2}),
yields
\begin{eqnarray}
\partial_t s_x + \partial_x j^x_x  & =  &- 2 m \alpha j^x_z - 
2 m
\beta j^y_z \\
\partial_t s_y + \partial_x j^x_y  &  =  & -2m\alpha j^y_z - 2m \beta
j^x_z \\
\partial_t s_z + \partial_x j^x_z  &  =  & 2 m \alpha (j^x_x + j^y_y )
+ 2m \beta (j_x^y + j_y^x)
\end{eqnarray}
where $j_\alpha^\beta$ is the spin current polarized in $\alpha$ and
flowing in $\beta $-direction.
In the diffusive limit the spin current densities are 
given by 
\begin{equation}
j_\alpha^\beta = - D {\partial_{ \beta} } s_\alpha + 2 \tau
\langle v_{F}^\beta ( {\bf b} \times {\bf s} )_\alpha \rangle
,\end{equation}
which allows to reproduce the spin diffusion equation, Eqs.\
(\ref{eq6})--(\ref{eq8}).
%
In narrow systems the slow modes have a smooth density profile, such
that to leading order in the system size the current can be considered
constant in space.
For a quantum dot, i.e.\ a system that is confined in all spatial
directions, the vanishing of the spin current through the
boundaries then implies immediately that $\partial_t {\bf s} = 0$.
For a narrow wire the situation becomes slightly more complicated
since only currents flowing into the boundary are zero, which after
some algebra leads to
\begin{equation}
\partial_t {\bf s} = -\frac{1}{\tau_s} \frac{1}{\alpha^2} 
\left(    
\begin{array}{ccc} 
\beta^2        &  \alpha \beta & 0 \\
\alpha \beta  &  \alpha^2      & 0 \\
0              &0               & \alpha^2 + \beta^2 
\end{array} \right) 
\left( \begin{array}{c} s_x \\ s_y \\ s_z  \end{array} \right)
.\end{equation}
In the absence of the Dresselhaus term ($\beta = 0$) this means that
$\partial_t s_x = 0$,  and $ \partial_t s_y =-s_y/\tau_s$, $\partial_t
s_z = - s_z /\tau_s$, which implies that  
the long-living mode in Fig.\ \ref{fig2} is polarized in $x$-direction.
In the presence of both a Rashba and a Dresselhaus term, the spin is
still conserved for one direction which depends 
on the relative strength of the two terms: Perpendicular to
the boundary when the Rashba term dominates, parallel to the  boundary when
the Dresselhaus term is larger, and somewhere in between (but
always in-plane) when both terms are comparable in size.

\begin{figure}
\includegraphics[width=0.42\textwidth]{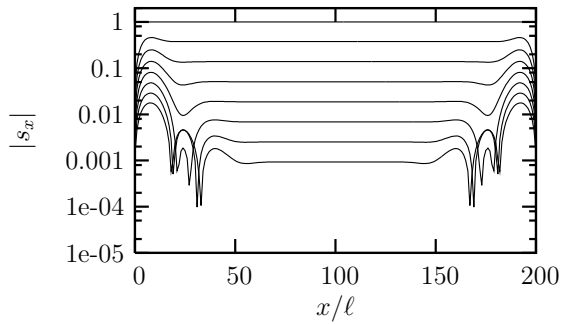}
\caption{\label{fig3}Time evolution of the spin polarization in
$x$-direction for a wire with
adiabatic boundary condition for the same set of parameters as in
Fig.~\ref{fig1}.}
\end{figure}

\begin{figure}
\includegraphics[width=0.42\textwidth]{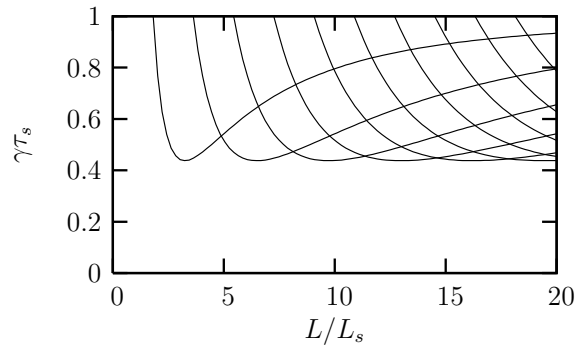}
\caption{\label{fig4}Lowest eigenvalues of the spin-diffusion operator for
adiabatic boundary conditions}
\end{figure}

The above results change considerably when different boundary conditions are
applied.
Fig.\ \ref{fig3} shows the time evolution of a spin polarization using  
adiabatic boundary conditions, Eq.~(\ref{eq26a}).
Here the spin has been prepared in $x$-direction, i.e.
perpendicular to the boundary.
In this case the boundary mode is absent, and the asymptotic decay of
the spin polarization is ruled by an inhomogeneous, but extended mode.
The spectrum of the spin diffusion operator is shown in Fig.\ \ref{fig4}.
The boundary condition implies that the eigenmodes
are
$s_{x,y} \propto \sin(q x)$, $s_z  \propto \cos( q x)$ with $q= n
\pi/L$;  the eigenvalues are given by 
\begin{equation} \label{eq26}
\gamma(q) = Dq^2 + \frac{3}{2 \tau_s} \pm \frac{1}{2\tau_s} \sqrt{1+
16L_s^2 q^2}. 
\end{equation}
By inserting the allowed $q$-values the spectrum shown in Fig.~\ref{fig4}
is reproduced. In contrast to the previous case of a spin-conserving boundary, 
here all the diffusion modes show an increasing spin relaxation rate at the smallest wire widths. 
In particular, all the modes show a non monotonous behavior as a function of the wire width with a minimum at 
$L/L_s =(4 \pi/\sqrt{15}) \ n \approx 3 \ n$, where $n$ is the mode index. 
For width $L<L_s$ all spin diffusion modes relax fast.

\section{Summary}
Motivated by a recent experiment\cite{holleitner2006} we studied spin relaxation in 
narrow wires with spin-orbit coupling. 
To this end we have solved the Eilenberger equation and the spin
diffusion equation in the presence of spin-orbit interaction\cite{raimondi2006}
supplemented by boundary conditions for both charge and spin degrees
of freedom assuming translational invariance along the wire.

The spin diffusion equation has extra,
off-diagonal gradient terms that couple the different spin
directions. Due to these terms there exist inhomogeneous
spin-profiles, typically spin-spirals, that
decay slower than a homogeneous spin configuration and that show up
close to the edges of the wire.
Therefore there is a tendency for the spin relaxation to slow down,
the stronger the influence of the boundary becomes, i.e.\
when decreasing the width of the wire. 

For very narrow wires, i.e.\ narrower than the spin relaxation length,
the spin relaxation rate depends crucially on the boundary condition
and on the form of the spin-orbit coupling.
For models with linear-in-momentum spin orbit field and spin-conserving
boundaries the spin relaxation rate goes to zero with the system size.
It is interesting to note that while in a quantum dot all
spin-components are conserved in the finite size limit, in the case of
a narrow wire only one of them is conserved. 
The direction of this conserved component depends on the relative strength of
the Rashba and Dresselhaus terms.

On the other hand
in the case of spin-active boundaries, the spin is not conserved 
and the relaxation rate grows when the wire becomes very narrow.
We believe that the experimentally observed\cite{holleitner2006} sharp 
increase of the spin relaxation rate in very narrow wires can be
explained by spin scattering at the boundary.

This work was supported by the Deutsche Forschungsgemeinschaft through
SFB484.


\begin{thebibliography}{999}
\bibitem{keldysh} More precisely, in a non-equilibrium Green function formalism, 
$g$ is the so-called Keldysh component  and may be seen as the quantum analog of the Boltzmann distribution function.
\end{thebibliography}
\end{document}